\documentclass[usenatbib]{mn2e}
\usepackage[letterpaper,margin=0.75in]{geometry}
\usepackage{natbibmnfix, graphicx, times, amsmath, epsfig, amssymb}
\usepackage{graphicx}
\usepackage{amsmath}
\usepackage{verbatim}

\newcommand{\Msun}{M_\odot}

\newcommand\lsim{\mathrel{\rlap{\lower4pt\hbox{\hskip1pt$\sim$}}
        \raise1pt\hbox{$<$}}}
\newcommand\gsim{\mathrel{\rlap{\lower4pt\hbox{\hskip1pt$\sim$}}
        \raise1pt\hbox{$>$}}}
\def\myputfigure#1#2#3#4#5%
{\vskip#5pt\makebox[0pt]{\hskip#2in
\includegraphics[width=#3\textwidth]{#1}}\vskip#4pt\hfill}

\newcommand{\apj}{ApJ}
\newcommand{\apjl}{ApJ}
\newcommand{\apjs}{ApJS}
\newcommand{\aap}{A\&A}

\newcommand{\mnras}{MNRAS}

\newcommand{\araa}{ARAA}

\begin{document}

\title[Gravitational Waves from Direct Collapse Black Holes Formation]
      {Gravitational Waves from Direct Collapse Black Holes Formation}
\author[F. Pacucci et al.]
{Fabio Pacucci$^1$,
Andrea Ferrara$^{1,2}$,
Stefania Marassi$^3$ \\
$^1$Scuola Normale Superiore, Piazza dei Cavalieri, 7  56126 Pisa, Italy \\
$^2$Kavli Institute for the Physics and Mathematics of the Universe (WPI), Todai Institutes for Advanced Study, the University of Tokyo \\
$^3$INAF/Osservatorio Astronomico di Roma, via di Frascati 33, 00040, Monteporzio, Italy \\
}
             
\date{submitted to MNRAS}

\maketitle
             
\begin{abstract}
The possible formation of Direct Collapse Black Holes (DCBHs) in the first metal-free atomic cooling halos at high redshifts ($z \gsim 10$) is nowadays object of intense study and several methods to prove their existence are currently under development.
The abrupt collapse of a massive ($\sim 10^4 - 10^5 \, \mathrm{M_{\odot}}$) and rotating object is a powerful source of gravitational waves emission.
In this work, we employ modern waveforms and the improved knowledge on the DCBHs formation rate to estimate the gravitational signal emitted by these sources at cosmological distances.
Their formation rate is very high ($\sim 10^4 \, \mathrm{yr^{-1}}$ up to $z\sim20$), but due to a short duration of the collapse event ($\sim 2-30\, \mathrm{s}$, depending on the DCBH mass) the integrated signal from these sources is characterized by a very low duty-cycle (${\cal D}\sim 10^{-3}$), i.e. a shot-noise signal.
Our results show that the estimated signal lies above the foreseen sensitivity of the Ultimate-DECIGO observatory in the frequency range $(0.8-300) \, \mathrm{mHz}$, with a peak amplitude $\Omega_{gw} = 1.1 \times 10^{-54}$ at $\nu_{max} = 0.9 \, \mathrm{mHz}$ and a peak Signal-to-Noise Ratio $\mathrm{SNR}\sim 22$ at $\nu = 20 \, \mathrm{mHz}$.
This amplitude is lower than the Galactic confusion noise, generated by binary systems of compact objects in the same frequency band. For this reason, advanced techniques will be required to separate this signal from background and foreground noise components. As a proof-of-concept, we conclude by proposing a simple method, based on the auto-correlation function, to recognize the presence of a ${\cal D} \ll 1$ signal buried into the continuous noise. \\
The aim of this work is to test the existence of a large population of high-z DCBHs, by observing the gravitational waves emitted during their infancy.
\end{abstract}


\setcounter{footnote}{1}

%%%%%%%%%%%%%%%%%%%%%%%%%%%%%%%%%%%%%%%%%%%%%%%%%%%%%%%%%%%%%%%%%%%%%%
%% SECTION 1: INTRODUCTION
%%%%%%%%%%%%%%%%%%%%%%%%%%%%%%%%%%%%%%%%%%%%%%%%%%%%%%%%%%%%%%%%%%%%%%
\section{Introduction}
\label{sec:intro}

Gravitational Wave (GW) observations are likely going to open a new window on the Universe within this decade. 
The different emitters of electromagnetic radiation and GWs are responsible for the dissimilar spatial regions they convey information about.
The electromagnetic radiation is emitted by charged particles with relatively small wavelengths, carrying information about the physical conditions of small spatial regions. On the contrary, the entire mass and momentum distributions of an astronomical source produce the emission of GWs, which carry information about large spatial regions, with very long wavelengths. Moreover, electromagnetic radiation and GWs couple very differently with the matter: while the former is easily absorbed or scattered, the coupling of the latter is extremely weak. This effect allows the GWs to travel cosmological distances without being affected, but also makes them very hard to be detected (see e.g. \citealt{Sathyaprakash_2009}).

First-generation GWs detectors (e.g. LIGO, Virgo) have been successfully
commissioned, and the development of the next advanced sensitivity ground-based detectors (Advanced LIGO, Advanced Virgo) is well underway. However, the construction of the planned space-based detectors (DECIGO, eLISA) is still in the development phase (see e.g. \citealt{DECIGO, eLISA_2013}).

The GW spectrum is conventionally divided into several regions, from extremely low ($\sim 10^{-18} \, \mathrm{Hz}$) to high ($\sim 10^{4} \, \mathrm{Hz}$) frequencies.  In the low frequency range (approximately from $10^{-4} \, \mathrm{Hz}$ to $1 \, \mathrm{Hz}$, see \citealt{Camp_2004} for an extensive review) the most important sources are Galactic compact binaries, massive Black Holes (BHs) binary mergers, massive BHs capture of compact objects, the collapse of super-massive stars and the primordial GWs background (see \citealt{Amaro-Seoane_2013} for an extensive review on the subject).

In particular, the collapse of Super-Massive Stars (SMSs), the first massive objects formed in unpolluted atomic cooling halos at $z \gsim 10$ may lead to a significant emission of GWs, depending on the collapse asymmetry. A possible mechanism for a highly asymmetrical collapse is the development of a dynamical bar-mode instability
as the super-massive star cools \citep{Smith_1996, Saijo_2002, Shapiro_2003, Franci_2013}. This may be likely if viscosity and magnetic fields are insufficient to keep the star rotating uniformly during the cooling phase. Given enough energy and a long enough lifetime of the bar mode, a significant fraction of the rest energy of the star could be lost as gravitational radiation (see e.g. \citealt{Schneider_2000}).
If this signal is actually observable by current or future GWs surveys, it would provide a highly valuable tool to prove the existence of Intermediate Mass Black Holes (IMBHs) at high redshifts.

In the commonly accepted $\Lambda$CDM cosmological scenario, the formation of the first stars and black holes occurred at $z = 20 - 30$ \citep{Miralda_E_2003, Bromm_Yoshida_2011,Volonteri_Bellovary_2012} in molecular cooling halos, i.e. dark matter structures with virial temperatures below $10^4 \, \mathrm{K}$ and masses below $M_{h} \sim 10^8 \, \mathrm{M}_{\odot}$.
These structures cooled their metal-free gas through molecular hydrogen line emission. 
Under intense UV irradiation, particularly in the Lyman-Werner (LW) band ($11.2-13.6 \, \mathrm{eV}$), the molecular hydrogen is photo-dissociated, so that the cooling is quenched (see \citealt{Machacek_2001,Fialkov_2012}).
On the contrary, larger metal-free halos (with masses above $\sim 10^8 \, \mathrm{M}_{\odot}$ and temperatures $T_{vir}>10^4 \, \mathrm{K}$), when illuminated by LW photons with fluxes above a threshold $J_{\nu}^{\bullet}$ (\citealt{Loeb_Rasio_1994,Eisenstein_Loeb_1995,Begelman_2006,Lodato_Natarajan_2006, Regan_2009, Shang_2010,Johnson_2012, Agarwal_2012, Latif_2013, Inayoshi_2012, Sugimura_2014, Dijkstra_2014}) are the ideal environment for the formation of Direct Collapse Black Holes (DCBHs).
Several theoretical works (\citealt{Bromm_Loeb_2003, Begelman_2006, Volonteri_2008, Shang_2010, Johnson_2012, Ferrara_2014}) have shown that the result of this collapse is the formation of a DCBH of mass $10^4 - 10^6\, \mathrm{M}_{\odot}$.
The collapse of SMSs is mainly driven by General Relativity (GR) instabilities, as described in \cite{Shibata_Shapiro_2002}. The subsequent accretion of mass contributes to the final mass of the IMBHs.

The existence of IMBHs at high-redshifts would provide a very convenient solution to two problems still unsolved by modern Cosmology, namely: (i) What are the sources responsible for the Cosmic Infrared Background (\citealt{Cappelluti_2013,Yue_2013})? (ii) How did the current Super-Massive Black Holes (SMBHs) form (see \citealt{Volonteri_Review_2012} for a recent review and references therein)?
We propose here a tentative test of the IMBH hypothesis based on the detection of the gravitational waves (GWs) emitted during the collapse phase.

Several works have been focused on the theoretical determination of the GWs waveform for the most important sources in the low frequency range, starting from the seminal paper \cite{Saenz_1978}. For example, \citealt{Schneider_2000, Ott_2004, Sekiguchi_2005, Ott_2009, Li_Benacquista_2010} have developed waveforms for the asymmetrical collapse of a SMS into a black hole both with simple theoretical arguments and, more recently, with numerical simulations.
For instance, \cite{Reisswig_2013} have investigated a non-axisymmetric instability which leads to the formation of a binary black hole system within the collapsing SMS.
Employing this modern waveform for the collapse of DCBHs at high redshifts ($z \sim 15$) we study their observability with the aid of recent estimates of their formation rates.

The outline of this paper is as follows. In $\S 2$ we present an overview of the theory describing the GWs emission, specifying the waveform used.
In $\S 3$ we describe the results and the possible observability of these gravitational signals.
Finally, in $\S 4$ we state our conclusions.
Throughout, we adopt recent Planck cosmological parameters \citep{Planck_Parameters_2013}: $(\Omega_m, \Omega_{\Lambda}, \Omega_b, h, n_s, \sigma_8 )= (0.32, 0.68, 0.05, 0.67, 0.96, 0.83)$,

%%%%%%%%%%%%%%%%%%%%%%%%%%%%%%%%%%%%%%%%%%%%%%%%%%%%%%%%%%%%%%%%%%%%%%
%% SECTION 2: THEORETICAL BACKGROUND
%%%%%%%%%%%%%%%%%%%%%%%%%%%%%%%%%%%%%%%%%%%%%%%%%%%%%%%%%%%%%%%%%%%%%%
\section{Theoretical Background}
\label{sec:theory}
The output of a GWs detector is a time series $s(t)$ which includes the instrument noise $n(t)$ and the response to the gravitational signal $h(t)$:
\begin{equation}
s(t) = P_+(t)h_+(t) + P_X(t)h_X(t) + n(t)
\end{equation}
The instrument response is a convolution of the antenna patterns $P_+(t)$ and $P_X(t)$ with the two GW polarizations $h_+$ and $h_X$. The antenna patterns depend on the sky location and on the emission frequency of the source and they are simple quadrupoles in the case of emission wavelengths which are large when compared to the detector baseline. 
The signal analysis is usually performed in the frequency domain since, in this representation, the noise is usually assumed to be uncorrelated and gaussian in each frequency bin.

The information contained in the time series is generally represented in the Fourier domain as a strain amplitude spectral density, $\tilde{h}(\nu)$. 
This quantity is defined in terms of the power spectral density:
\begin{equation}
S_s(\nu) = \tilde{s}^*(\nu) \tilde{s}(\nu)
\end{equation}
The tilde operator is the Fourier transform of the time series:
\begin{equation}
\tilde{s}(\nu) = \int_{-\infty}^{+\infty} \! s(t) e^{2 \pi i \nu t} \, \mathrm{d}t.
\end{equation}
and the star indicates the conjugation operation in complex numbers.
The power spectral density $S_s(\nu)$ has units of time or, as usually indicated, $\mathrm{Hz}^{-1}$.
The strain amplitude spectral density is then defined as:
\begin{equation}
\tilde{h}(\nu) = \sqrt{S_s(\nu)}
\end{equation}
This quantity has units $\mathrm{Hz}^{-1/2}$ and is the one commonly reported in plots showing gravitational signals.
From the prediction of the time evolution of the gravitational signal shape $h(t)$, derived from the instrumental time series $s(t)$, it is then possible to compute the strain amplitude spectral density $\tilde{h}(\nu)$.

It is standard practice to quote the strength of a gravitational signal in terms of the energy density per logarithmic frequency interval, $d\rho_{gw}/d\mathrm{ln}\nu$, scaled by the energy density needed to close the Universe:
\begin{equation}
\Omega_{gw}(\nu) = \frac{1}{\rho_{crit}} \, \frac{d\rho_{gw}}{d \mathrm{ln} \nu} = \frac{4 \pi^2 }{3 H_0^2} \nu^3 S_s(\nu)
\end{equation}
where $H_0$ is the Hubble constant, $\rho_{crit}$ is the critical density and $\rho_{gw}$ is the energy density in a GW, given by:
\begin{equation}
\rho_{gw} = \frac{1}{32 \pi G} \, <h^2>
\end{equation}
where $G$ is the gravitational constant and the brackets denote a spatial average over the wavelengths.

The actual prediction of the shape $h(t)$ is far from being trivial and strongly depends on the nature of the source and on its physical properties. Only recently a number of accurate waveforms have been proposed for binary systems and for collapsing SMSs (see the works \citealt{Schneider_2000, Li_Benacquista_2010, Li_Koksis_Loeb_2012, Ajith_2011, Pan_2014}).
The form of the gravitational waveform template plays a crucial role for data analysis: very often, only a detailed prediction of the signal shape allows to disentangle it from the receiver noise and/or from other sources of gravitational radiation. We will discuss this point more carefully in the following subsections.

\subsection{Waveform for IMBH Collapse and Ringdown}
A DCBH can emit GWs at its formation through the asymmetric core collapse of its progenitor.
Indeed, in order to emit gravitational radiation, a physical system needs to be non-spherical: one possibility for having a non-spherical collapse is requesting the primordial halo, hosting the to-be-formed BH, to be in rotation.
DCBHs are formed during a brief era of cosmic time ($ 13 \lsim z \lsim 20$, \citealt{Yue_2014}) due to the collapse of the inner part of metal-free halos.
The angular momentum of high-redshift halos is small \citep{Davis_2010} and is preserved during the collapse. The newly formed DCBH is then characterized by low values of angular momentum, which is later increased by accretion and merging events.

We focus on BH progenitors, i.e. the inner section of the collapsing halo, with mass $M$, mean mass density $\rho$ and dimensionless spin parameter $a$ defined as:
\begin{equation}
a = \frac{Jc}{GM^2}
\end{equation}
where $J$ is the angular momentum and $c$ is the speed of light. Note that $a=0$ denotes a non-rotating BH, while $a=1$ is a maximally rotating Kerr BH, although we note that \cite{Thorne_1974} showed that accretion driven spin is limited to $a = 0.998$. Magnetic fields connecting material in the disk and the plunging region may further reduce the equilibrium spin: magneto-hydrodynamic simulations for a series of thick accretion disks suggest an asymptotic equilibrium spin at $a \sim 0.9$ \citep{Gammie_2004}.
Rotating BH progenitors will distort to an oblate spheroid shape despite their own immense gravity. The collapse, under this non-spherical geometry, causes the emission of a huge amount of energy through GWs.
The final evolution of the collapsing halo is usually characterized by two distinct phases:
\begin{enumerate}
\item \textbf{Collapse:} under the influence of its overwhelming self-gravity, the BH progenitor collapses in a time scale equal to the dynamical time $t_{dyn} \sim 1/ \sqrt{G\rho}$.
\item \textbf{Ringdown:} the collapsed non-spherical object undergoes strong oscillations which progressively allow it to acquire a spherical shape.
\end{enumerate}
As we shall see, the maximum of gravitational radiation is released at the end of the collapse, when the BH bounces back and starts the subsequent ringdown phase.

In the work by \cite{Li_Benacquista_2010}, the core collapse of the SMS is approximated as an axisymmetric Newtonian free-fall of a rotating relativistic degenerate iron core. In addition, the collapse waveform is reasonably well modeled by an exponential growth.
Following this paper, the time-series of the waveform for the ringdown phase is expressed analytically as:

\begin{equation}
h_{RD}(t) = A\frac{G M}{c^2 d_L} \mathrm{exp}({-\pi \nu_0 t/Q}) \cos({2 \pi \nu_0 t})
\end{equation}
where $d_L$ is the luminosity distance between the source and the detector, $Q = 2(1-a)^{-9/20}$, $g(a) = 1 - 0.63(1-a)^{3/10}$, $\nu_0 = c^3 g(a) / (2 \pi G M)$.
Note that $\nu_0^{-1}$ is the typical timescale of GWs emission.
In addition:
\begin{equation}
A = \sqrt{\frac{5 \epsilon}{2}} Q^{-1/2} \left(1+\frac{7}{24Q^2}\right) g(a)^{-1/2}
\end{equation}
where $\epsilon$ is the fraction of the initial mass of the BH which is transformed into GWs radiation. Numerical simulations have shown that approximately 1\% of the final BH mass is emitted in GWs so, throughout, we adopt the value $\epsilon=0.01=1\%$ (see \citealt{Buonanno_2007, Abbott_2009} for details). 
In the latter work, the authors investigate the presence of possible GWs burst signals in the high-frequency range $1-6 \, \mathrm{kHz}$, without finding any evidence of them, but putting an indirect upper limit on the emitted energy.
Although the survey was dedicated to a different frequency range, we make use of the same upper limit, in accordance with \cite{Li_Benacquista_2010}.

\begin{center}
\begin{figure}
\vspace{-1\baselineskip}
\hspace{-0.5cm}
\includegraphics[angle=0,width=0.48\textwidth]{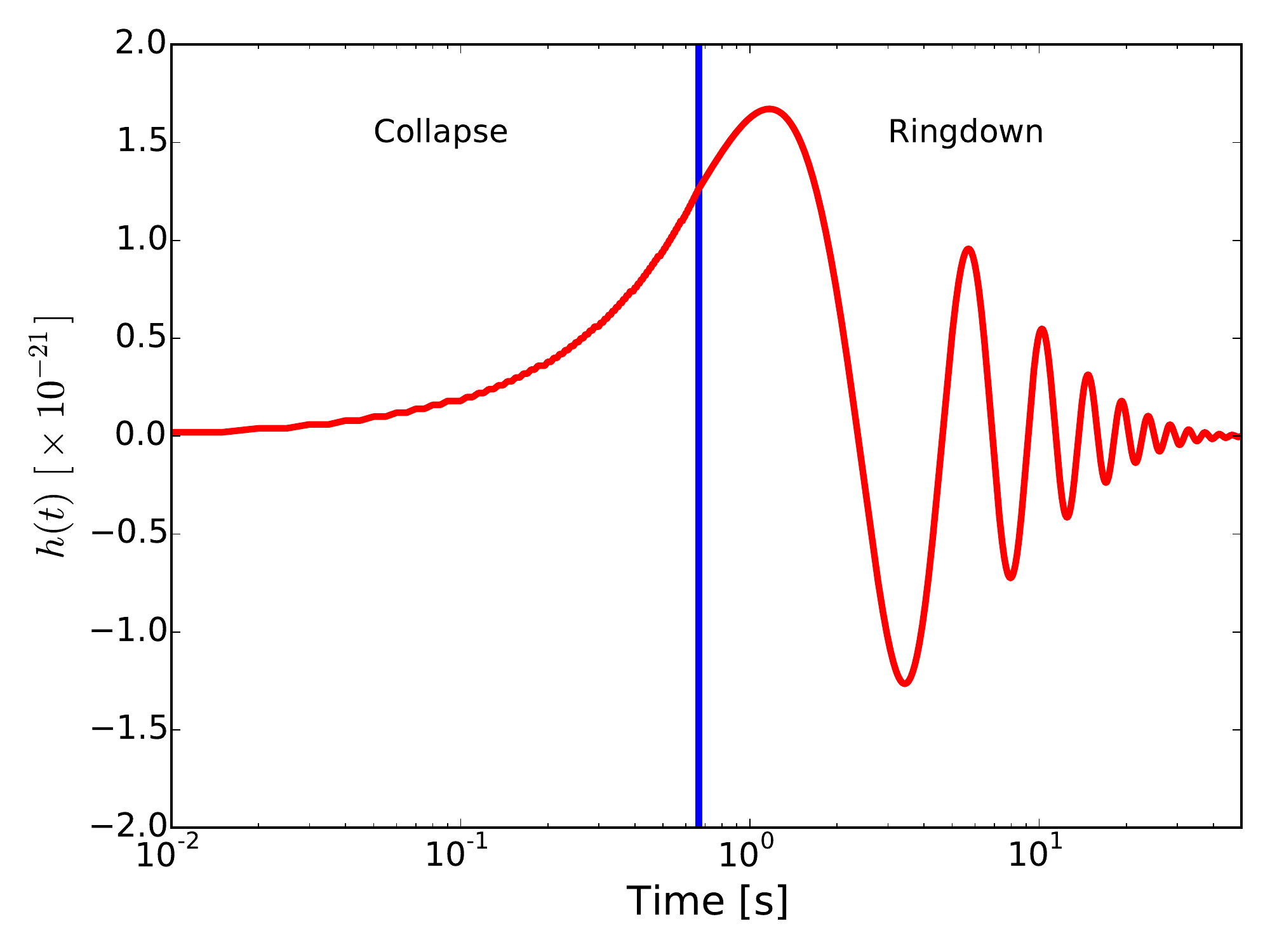}
\caption{Gravitational strain as a function of time for the collapse of a $10^5 \, \mathrm{M_{\odot}}$ BH at $z=15$, with a mean mass density $\rho=10^7 \, \mathrm{g\, cm^{-3}}$. The separation between the collapse and the ringdown phases is marked by the vertical blue line.}
\label{fig:time_series_example}
\end{figure}
\end{center}

\begin{center}
\begin{figure}
\vspace{-1\baselineskip}
\hspace{-0.5cm}
\includegraphics[angle=0,width=0.48\textwidth]{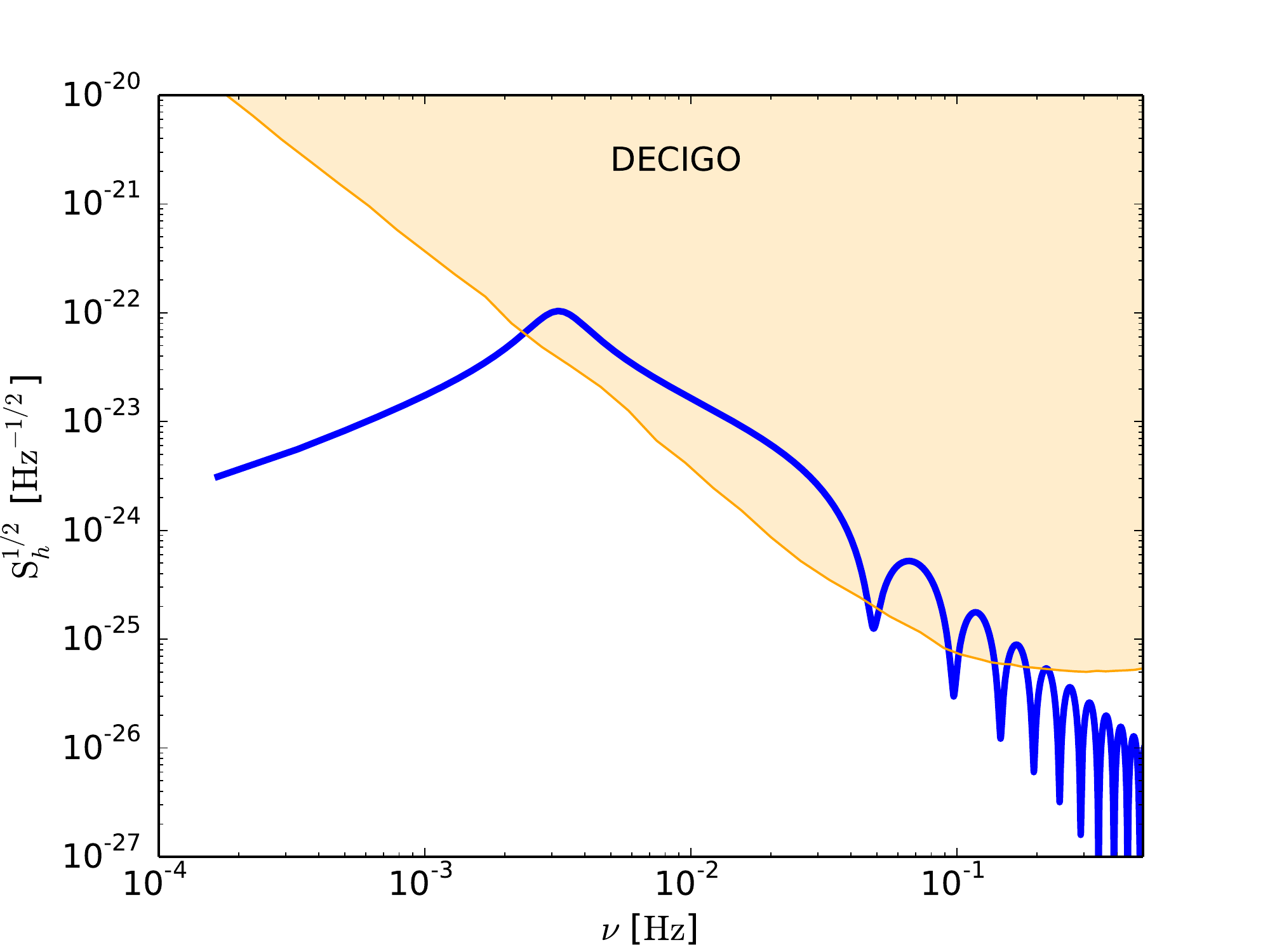}
\caption{Spectral strain for a single source, due to the collapse of a $10^{5} \, \mathrm{M_{\odot}}$ BH at $z=14$, with a mean mass density of $\rho=10^7 \, \mathrm{g\, cm^{-3}}$. The power peak is reached at $\nu \sim 3 \times 10^{-3}\, \mathrm{Hz}$. The sensitivity curve for the DECIGO observatory is reported as a shaded orange line.}
\label{fig:spectral_strain_example}
\end{figure}
\end{center}

The waveform for the collapse is found by analytically fitting a numerical relativistic simulation in \cite{Li_Benacquista_2010}. The time series has the following general form:
\begin{equation}
h_C(t) \sim \alpha + \frac{a}{\rho}e^{\gamma M t}
\end{equation}
where $a$ is the spin parameter, $M$ is the mass, $\rho$ is the mean mass density of the collapsing object, and $\alpha$ and $\gamma$ are two free parameters that we determine by imposing that $h_C(t=0)=0$ and $h_C(t=t_c) = h_{RD}(t=t_c)$, i.e. the gravitational signal is null at the initial time and the two waveforms match at the collapse time $t_c$.
The collapse time is found by imposing its identity with the dynamical time of a quasi-spherical object with a mean mass density $\rho$:
\begin{equation}
t_c = t_{dyn} \sim \frac{1}{\sqrt{G \rho}}
\end{equation}
Calling $V$ the value of the ringdown waveform at the collapse time $t_c$, our final form for the collapse waveform is the following:
\begin{equation}
h_C(t) = \frac{a}{\rho}\left(e^{\gamma M t}-1\right)
\label{collapse_waveform}
\end{equation}
with
\begin{equation}
\gamma = \frac{1}{M t_c} \mathrm{ln}\left(\frac{\rho}{a}V+1 \right)
\end{equation}
The physical parameters in the model are: the mean mass density of the collapsing object $\rho$, the BH mass $M$ and the spin parameter $a$. The range of their values used in our calculations is detailed in Sec. \ref{sec:results}.
Fig. \ref{fig:time_series_example} shows an example of time series for the gravitational strain: the collapse and subsequent ringdown phases are evident and separated by the vertical line at $t_c \sim t_{dyn} \sim 0.7 \, \mathrm{s}$.
Fig. \ref{fig:spectral_strain_example} shows the spectral strain for a single source with the parameters reported in the caption. The peak amplitude $\sim 10^{-22} \, \mathrm{Hz^{-1/2}}$ is reached at $\nu \sim 3 \times 10^{-3}\, \mathrm{Hz}$. 

Other waveforms have been proposed for the collapse of SMSs with lower masses. For example, \cite{Suwa_2007} employed GR numerical simulations and the standard quadrupole formula developed in \cite{Moenchmeyer_1991} to study the collapse of $\sim 500 \, \mathrm{M_{\odot}}$ SMSs, obtaining a peak power in the range $10-100 \, \mathrm{Hz}$. Due to important differences in the collapse modeling, a direct comparison between the classical quadrupole formula derived in \cite{Moenchmeyer_1991} and the one used in the present work is hard to obtain.

\subsection{Other Sources of Gravitational Signals in the Low Frequency Band}
Most of the gravitational power emitted from the DCBH collapse falls in the low frequency range, approximately from $10^{-4} \, \mathrm{Hz}$ to $1 \, \mathrm{Hz}$ (see \citealt{Camp_2004} for an extensive review).
In this range the most important sources of GWs are Galactic and extragalactic compact binaries, massive BHs binary mergers, massive BHs capture of compact objects and the primordial GW background.
The last one is too weak to be of any concern in this work (see e.g. \citealt{Yagi_2011}). 
This is not the case with the Galactic and extragalactic compact binaries.

The unresolved background produced by the GWs emission of Galactic and extragalactic compact binaries acts as a confusion noise in this frequency range \citep{Schneider_2001, Farmer_2003, Nelemans_2009, Sathyaprakash_2009,Regimbau_2010, Marassi_2011, Regimbau_2012}.
This is produced by the coalescence of two neutron stars (NS-NS), two black holes (BH-BH) or a neutron star and a black hole (NS-BH) in our Galaxy or in the $z\lsim5$ Universe. The duty-cycle (i.e. the percentage of one period in which a signal is active, see a more complete definition below) for these events, for future space-borne gravitational observatories like DECIGO, will be higher than unity, i.e. the signals overlap and create a confusion noise of unresolved sources.

The merging event for SMBHs also produces very energetic GWs \citep{Kocsis_Loeb_2008} that may represent an important foreground contribution to the detected gravitational signal. One very simple method to reproduce the gravitational signals generated by SMBHs merging events is presented in \cite{Li_Koksis_Loeb_2012} where the interested reader is referenced to for a detailed description. However, the signal generated by these events is very different (in terms of both the duty-cycle and the characteristic energy) from the one of our interest and their separation should be simple. 

\subsection{Formation Rates of DCBHs}
We employ the mass density of high-redshift DCBHs predicted in the work \cite{Yue_2013}, where they present the rate of formation per unit volume of DCBHs, $\dot{\varphi}(z)$, that accounts for the observed Cosmic Infrared Background (CIB) fluctuation excess. The Universe enters the DCBHs era at $z \approx 20$ when a large fraction of atomic cooling haloes are experiencing DCBHs formation. Their formation is suppressed after $z \approx 13$, so that the DCBHs era lasts only $\sim 150 \, \mathrm{Myr}$ of the cosmic history (see \citealt{Yue_2014} for details).
Then, the number of GWs sources formed per unit time out to a given redshift $z$ can be computed by integrating the cosmic DCBHs formation rate per unit volume $\dot{\varphi}(z)$, see Fig. \ref{fig:DC_rate_ensemble}.

It is important to note that the \cite{Yue_2013, Yue_2014} estimated DCBH high-$z$ mass density ($\sim 2 \times 10^6 \, \mathrm{\Msun \, Mpc^{-3}}$) exceeds the present-day SMBH mass density ($\sim 2 \times 10^5 \, \mathrm{\Msun \, Mpc^{-3}}$, \citealt{Yu_2002}) computed through the standard Soltan argument (which derives this quantity from the quasar output, see \citealt{Soltan_1982} for the original paper). This point is worth some discussion.

First, the \cite{Yue_2013, Yue_2014} DCBH abundance is required in the extreme case in which most of the CIB fluctuations in the range $(3.6 - 4.5) \,  \mathrm{\mu m}$ are to be explained by DCBHs. This might not be the case, and the difference between the local BH density and the high-$z$ DCBH one might be eased. 

Second, even if the CIB constraint is imposed, the BH mass density adopted here cannot be ruled out by current observations.
In fact, estimates derived from SMBH demographics refer to BHs that (i) are relatively massive, and (ii) reside in structures that are identified as ``galaxies''. As for point (i), there could be a large number of BHs with masses $\sim 10^{4-5} \, \mathrm{\Msun}$ that are too faint to contribute sensibly to the X-Ray Background (XRB) and/or are very inefficiently accreting (see e.g. studies considering slim-disk solutions like \citealt{Begelman_1982, Paczynski_1982, Sadowski_2009, Sadowski_2011}, and the recent work by  \citealt{Volonteri_2014}). These objects could be easily misidentified with other X-ray sources, e.g. High-Mass X-Ray Binaries (HMXBs) and Ultra-Luminous X-ray sources (ULXs). The additional possibility (ii) exists that a considerable fraction of DCBHs resides in baryon-free dark matter halos, as suggested by recent studies, e.g. \cite{Pacucci_2015}. These authors have simulated the accretion process onto a DCBH of initial mass $10^5 \, \mathrm{\Msun}$ at $z \sim 10$, embedded in a dark matter halo with a gas content of $\sim 10^7 \, \mathrm{\Msun}$, finding that after $\sim 142 \, \mathrm{Myr}$ about $90\%$ of the initial gas mass of the halo is accreted onto the compact object. If these objects are not embedded in larger gaseous structures they could completely escape our census.

Finally, \cite{Kormendy_2013} suggested a revision of the BH mass scaling relations, indicating that the local mass density in black holes should be increased by up to a factor of five with respect to previously determined values. As already pointed out, a possibility to explain a large BH mass density is that most of their growth occurs via radiatively-inefficient channels. In a recent paper, \cite{Comastri_2015} showed that it is possible to accommodate a large fraction of heavily buried, Compton-thick AGNs, without violating the limit imposed by the hard X-ray and mid-infrared backgrounds spectral energy density.

Thus, before we can firmly exclude the existence of a substantial population of intermediate mass black holes at high-$z$ one would need to predict in detail where these DCBHs end up locally, what is their accretion history, and what are the best strategies to detect them. The aim of this work is to explore the detectability of these objects via their gravitational wave signal, at least in the optimistic, but physically motivated, case considered here.

\subsection{Calculation of Event Rates, Duty-Cycles and S/N Ratios}
An important parameter to describe the signal in the time domain is the duty-cycle, ${\cal D}$. This is defined as the ratio between the typical duration of the signal emitted by a single source and the average time interval between two successive emissions.
When ${\cal D} \gsim 1$, the overall signal is continuous (i.e. the different signals overlap in time), conversely if ${\cal D}<1$ the resulting background is characterized by a shot-noise structure.
From the value of $\dot{\varphi}(z)$ and from the Initial Mass Function (IMF) $\Phi(M)$ of the population, it is possible to compute the duty-cycle as:
\begin{equation}
\frac{d{\cal D}(z)}{dz}=\frac{\dot{\varphi}(z)}{(1+z)} \frac{dV}{dz} \frac{(1+z)}{\nu_0} \, \int \! \Phi(M) \, \mathrm{d}M.
\end{equation}
Here, the IMF of IMBHs seeds (normalized by the mass) is taken from \cite{Ferrara_2014} and $\nu_0^{-1}$ is the typical timescale of GWs emission (see Sec. \ref{sec:theory}). The double appearance of the term $(1+z)$ is to take into account the redshift dependence of both $\nu_0$ and $\dot{\varphi}$.

\begin{center}
\begin{figure}
\vspace{-1\baselineskip}
\hspace{-0.5cm}
\includegraphics[angle=0,width=0.48\textwidth]{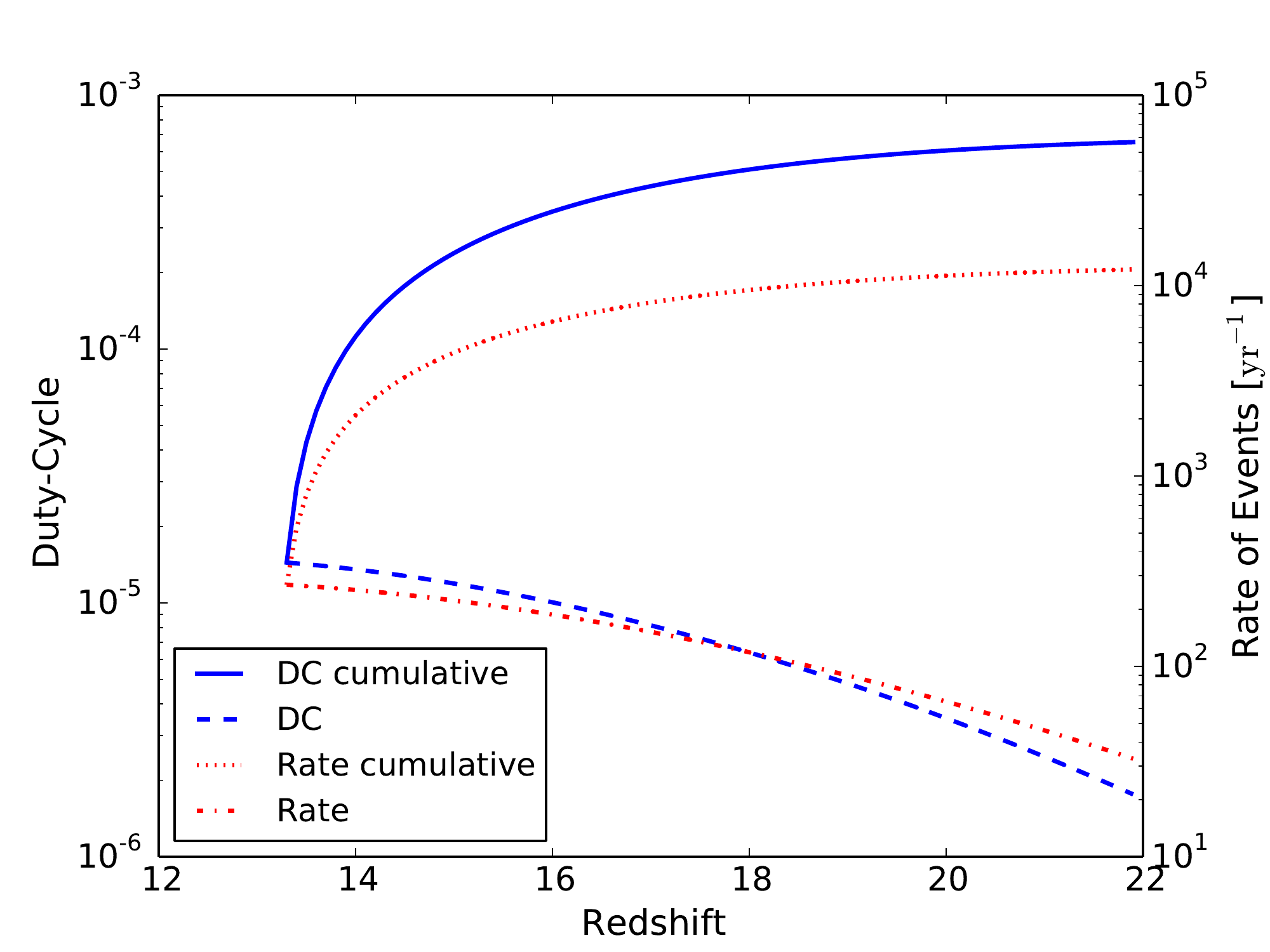}
\caption{The event rate and the duty-cycle for the formation of DCBHs as a function of redshift. The upper lines (solid and dotted ones) are cumulative, i.e. obtained summing all the values up to the given redshift.}
\label{fig:DC_rate_ensemble}
\end{figure}
\end{center}

The event rate and the duty-cycle are shown in Fig. \ref{fig:DC_rate_ensemble}.
The event rate for all-sky observations is very high, reaching a peak of $\sim 10^4 \, \mathrm{yr^{-1}}$ at $z\sim22$. However, due to a rather short duration of the GW burst ($\sim 2-30\, \mathrm{s}$ depending on the DCBH mass, see Fig. \ref{fig:time_series_example}) the duty-cycle is very small, of order $10^{-3}$. This result suggests that the gravitational signal produced by the collapse of high-redshift DCBHs is a shot-noise. The decrease, with increasing redshift, of both the event rate and the duty-cycle (and the corresponding leveling of their cumulative values) is ascribed to the decrease in $\dot{\rho}(z)$, as in \cite{Yue_2013}.
Both the event rate and the duty-cycle are comparable with the results in \cite{Marassi_2009} who, however, considered the contribution to the stochastic background of Pop III stars with masses in the range $(300 - 500) \, \mathrm{M}_{\odot}$ that collapse to BHs. 

The Signal-to-Noise Ratio allows a direct comparison between the GW amplitude and the instrumental sensitivity, assessing what sort of sources will be observable against noise.
A stationary noise is usually described by its Power Spectral Density (PSD): as a consequence, these comparisons are most commonly made in the frequency domain.

In this work, we deal with short-lived signals, which have wide bandwidths and the actual observation time of the source is not relevant in the determination of the Signal-to-Noise Ratio (SNR).
It is useful to define the dimensionless noise power per logarithmic bandwidth:
\begin{equation}
\hat{h}_n^2(\nu) = \nu S_n(\nu)
\end{equation}
where $\hat{h}_n(\nu)$ is called the effective GW noise.
From the signal side, we may define the characteristic signal amplitude:
\begin{equation}
\hat{h}_c = \nu |\tilde{h}(\nu)|
\end{equation}
which is dimensionless. 
This last quantity is to be compared with the effective GW noise $h_n(\nu)$ to obtain a rough estimate of the SNR of the signal: 
\begin{equation}
SNR(\nu) = \frac{h_c(\nu)}{h_n(\nu)}
\end{equation}
The sensitivity curve for Ultimate-DECIGO has been taken from \cite{Marassi_2009} and \cite{Yagi_2011}.

The SNR with respect to the instrumental noise is not the only parameter to be taken into account in order to assess the observability of a source. The signal also needs to be distinguished from all the other components received in the same frequency band, as described previously in this Section. The observability will be addressed more carefully in Sec. \ref{sec:results}.

%%%%%%%%%%%%%%%%%%%%%%%%%%%%%%%%%%%%%%%%%%%%%%%%%%%%%%%%%%%%%%%%%%%%%%
%% SECTION 3: RESULTS
%%%%%%%%%%%%%%%%%%%%%%%%%%%%%%%%%%%%%%%%%%%%%%%%%%%%%%%%%%%%%%%%%%%%%%
\section{Results}
\label{sec:results}
In this Section we present the results of our calculations and we discuss on the observability of the signal.
In Sec. \ref{sec:theory} we presented the employed waveform for the gravitational signal generated by the collapse of a high-redshift atomic cooling halo, with the subsequent formation of a DCBH. The waveform is a function of the mean mass density $\rho$, the total mass $M$ and the spin parameter $a$ (see Eq. \ref{collapse_waveform}). The amplitude of the detected gravitational signal depends also on the luminosity distance from the source, i.e. from its redshift.

In order to simulate the gravitational signal generated by a realistic ensemble of the population of high-z DCBHs, we let these parameters vary into different ranges of values, summarized in the following Table 1.

\begin{center}
\begin{tabular}{c | c | c | c}
\hline
  & MIN & MAX & Bins\\
\hline
$z$  & 14 & 22 & 10 \\
$\mathrm{Log} \, \rho \, [\mathrm{g\,cm^{-3}}]$  & 6    & 8 & 10 \\
$\mathrm{Log} \, M \, [\mathrm{M_{\odot}}]$ & 4.5 & 5.7 & 10\\
\hline
\end{tabular}
\label{tbl:ranges}
\end{center}
The redshift range has been chosen in accordance with the results of \cite{Yue_2013}, while the mass range to be compatible with the IMF values in \cite{Ferrara_2014}.
The density interval has been taken from simulations in \cite{Li_Benacquista_2010}, while the variation of the spin parameter $a$ for high-z BHs from \cite{Volonteri_2010}, \cite{Volonteri_2013}. Generally speaking, the rotational energy of primordial BHs is very low compared with local ones: the spin is progressively increased by the accretion of matter onto the compact object and by merging events. For this reason, our average value for the spin parameter is $a \sim 0.05$ (see also \citealt{Davis_2010}).

The variation of these physical quantities have different impacts on the produced gravitational signal.
A study of their effect has been sketched in the following Fig. \ref{fig:Comp_mass_z} and Fig. \ref{fig:Comp_spin_rho}.
In these figures the range of variation for the different parameters is restricted to the values shown in the legend.

\begin{center}
\begin{figure}
\vspace{-1\baselineskip}
\hspace{-0.5cm}
\includegraphics[angle=0,width=0.48\textwidth]{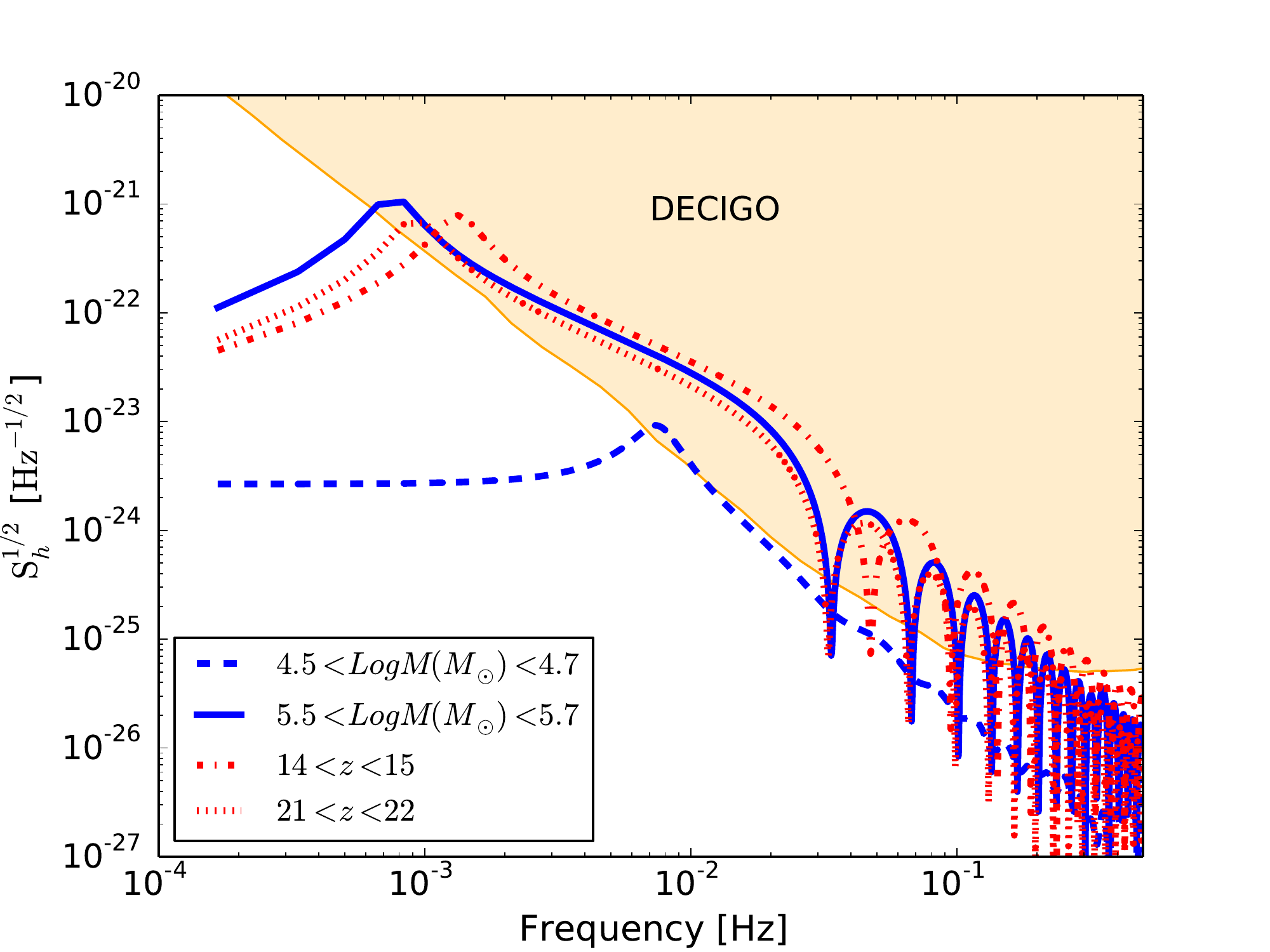}
\caption{Effect on the strain amplitude $\tilde{h}(\nu)$ of the BH mass and of its redshift. In this case the calculation has been performed in narrow ranges of the parameters, as indicated in the legend. The orange region is the sensitivity for the future Ultimate-DECIGO observatory.}
\label{fig:Comp_mass_z}
\end{figure}
\end{center}

\begin{center}
\begin{figure}
\vspace{-1\baselineskip}
\hspace{-0.5cm}
\includegraphics[angle=0,width=0.48\textwidth]{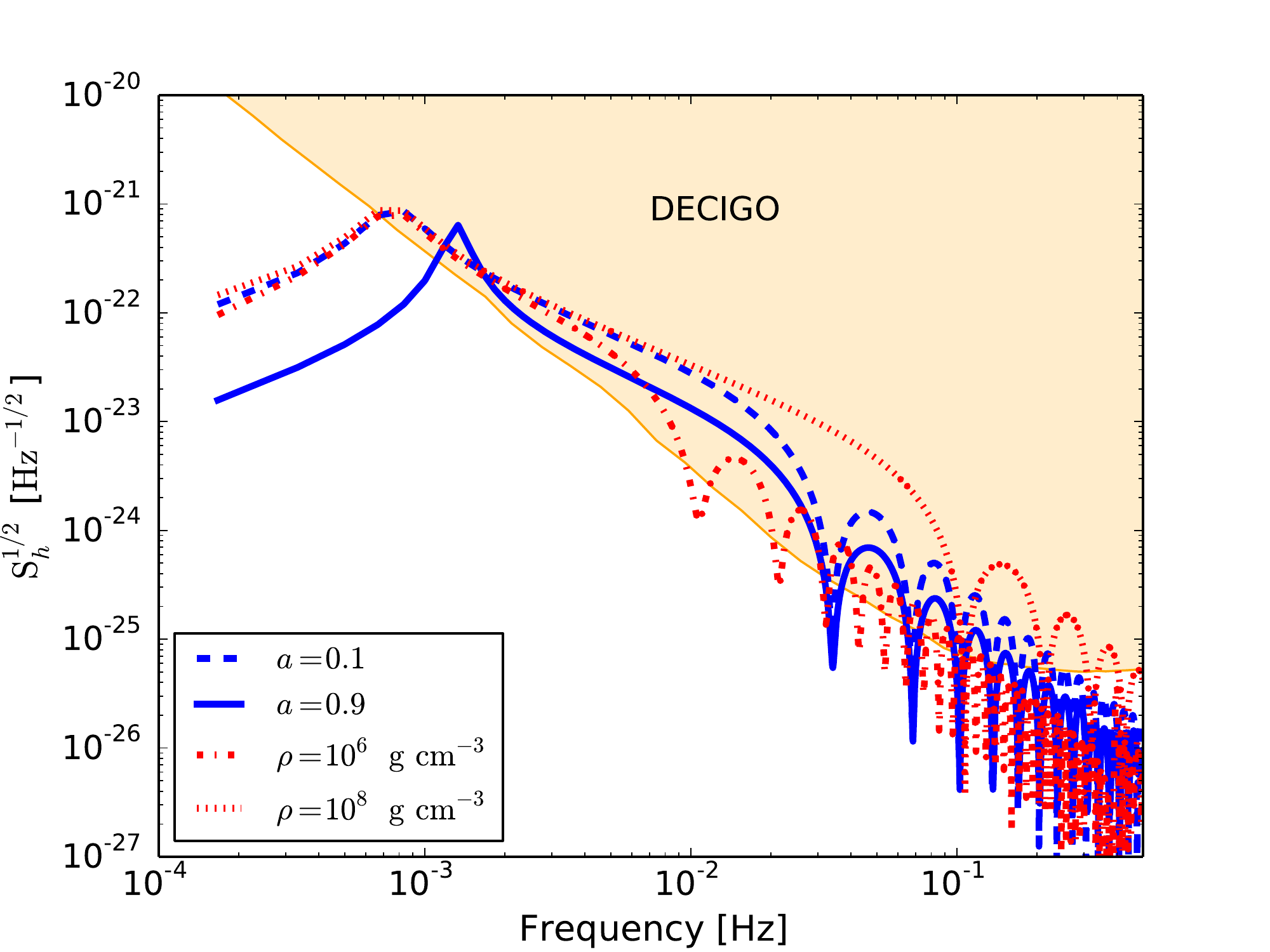}
\caption{As in Fig. \ref{fig:Comp_mass_z}, but for the spin parameter and the mean density. In this case, the single values used for the parameters are shown in the legend.}
\label{fig:Comp_spin_rho}
\end{figure}
\end{center}

The effect of the redshift on the gravitational signal is very small. This is due to the fact that the whole range of redshift values spans only $\sim 150 \, \mathrm{Myr}$ of cosmic history. For higher redshifts, the observed gravitational signal is shifted to lower frequencies, as expected. On the contrary, the effect of the mass is greatly important: lower masses correspond to higher frequencies of the peak ($\nu_0 \propto 1/M$, see Sec. \ref{sec:theory}), but also to an overall amplitude of the signals $\sim 2$ orders of magnitude lower, at the extreme detection limit with the DECIGO observatory. The spin parameter affects mainly the shape of the spectral strain amplitude and its peak frequency, which is shifted to higher values for larger values of $a$, while the amplitude is not affected. The mean mass density $\rho$ has virtually no effect on the peak frequency and on its amplitude, but varies the shape of the spectrum at higher frequencies. This is due to the fact that the mean mass density directly controls the duration of the collapse phase, $t_c \sim t_{dyn} \sim 1/\sqrt{G\rho}$, i.e. the transition from the collapse to the ringdown phases.

\begin{center}
\begin{figure}
\vspace{-1\baselineskip}
\hspace{-0.5cm}
\includegraphics[angle=0,width=0.48\textwidth]{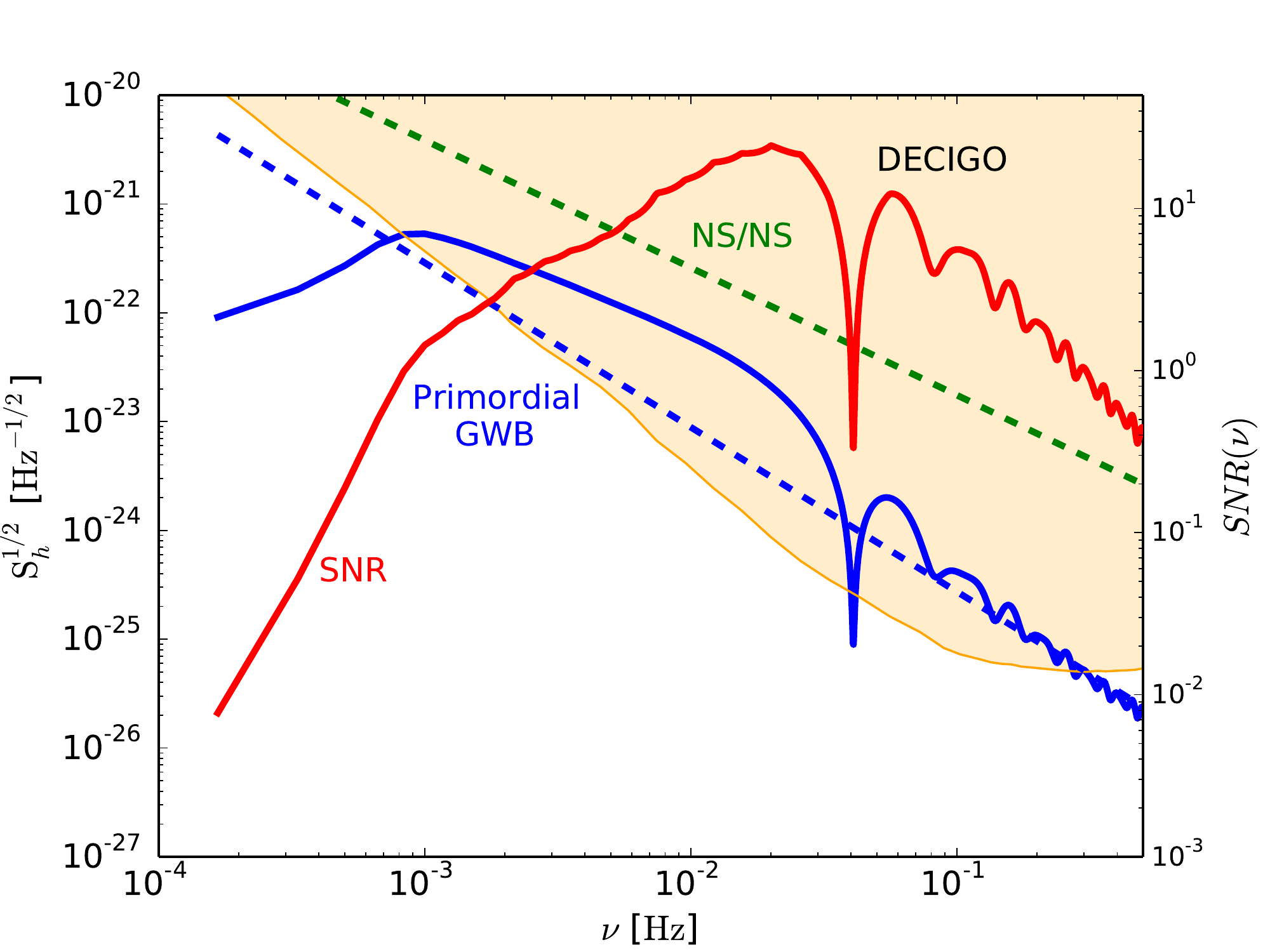}
\caption{Gravitational strain $\tilde{h}(\nu)$ for the total ensemble. The orange region is the sensitivity for the future Ultimate-DECIGO observatory. The dashed lines indicate the Galactic Confusion Background (GCB) due to the GWs emission from double neutron stars (NS/NS), which is the dominant foreground component (see \citealt{Yagi_2011}).}
\label{fig:strain_ensemble}
\end{figure}
\end{center}

The gravitational strain $\tilde{h}(\nu)$ for the total ensemble is reported in Fig. \ref{fig:strain_ensemble}.
The orange region is the sensitivity curve of Ultimate-DECIGO, as obtained from \cite{Marassi_2009} and \cite{Yagi_2011}, and it has been used to compute the SNR, reported as a solid red line in Fig. \ref{fig:strain_ensemble}.
The estimated signal lies above the foreseen sensitivity of Ultimate-DECIGO in the frequency range $(0.8-300) \, \mathrm{mHz}$, with a peak amplitude $\Omega_{gw} = 1.1 \times 10^{-54}$ at $\nu_{max} = 0.9 \, \mathrm{mHz}$.
The Signal-to-Noise ratio reaches a maximum value of $\mathrm{SNR}\sim 22$ at $\nu = 20 \, \mathrm{mHz}$. A source is considered detectable if the resulting SNR exceeds some standard threshold value, typically between 5 and 10 \citep{Plowman_2010}, so the signal generated by the collapse of high-redshift atomic cooling halos into DCBHs is detectable by the future Ultimate-DECIGO observatory.

However, in this frequency interval the sensitivity of Ultimate-DECIGO is limited by the unresolved background produced by the GW emission of Galactic compact binaries that acts as a confusion noise \citep{Nelemans_2009, Regimbau_2010, Regimbau_2012}: this is shown in Fig. \ref{fig:strain_ensemble} with a green dashed line \citep{Sathyaprakash_2009}. The duty-cycle for these events, for future space-borne gravitational observatories like DECIGO, will be higher than unity, i.e. the signals overlap and create a confusion noise of unresolved sources. See \cite{Yagi_2011} for a detailed description of the binary confusion noise as observed by the future DECIGO observatory.

\begin{center}
\begin{figure}
\vspace{-1\baselineskip}
\hspace{-0.5cm}
\includegraphics[angle=0,width=0.48\textwidth]{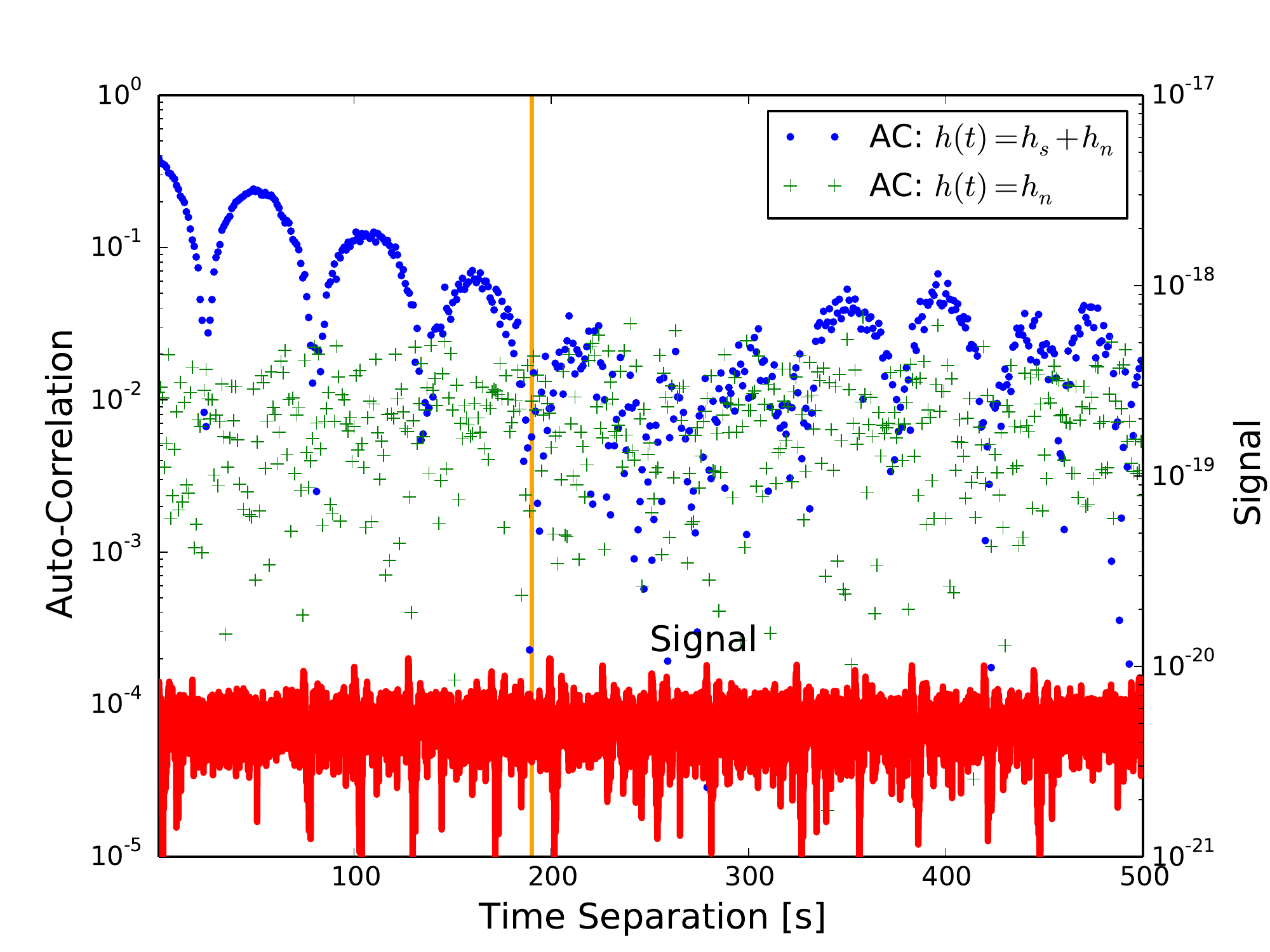}
\caption{Simple example of a signal processing method to separate a signal with ${\cal D} \ll 1$ from the confusion noise. The sum $h(t) = h_s(t) + h_n(t)$, where $h_s(t)$ is the shot-noise signal and $h_n(t)$ is the background noise component, is shown on the bottom, in red. The blue circles and the green crosses are the auto-correlation function for $h(t)$ and for $h_n(t)$ only, respectively. The duty-cycle of the signal is ${\cal D}=0.2$ (this value has been chosen for an easier visualization on the plot), while the population of collapsing DCBHs has a realistic distribution of masses ($10^{4} \, \mathrm{\Msun}<M<10^{5} \, \mathrm{\Msun} $, spin paramenters ($0.01<a<0.5$) and densities ($10^6 \, \mathrm{g \, cm^{-3}}<\rho<10^8 \, \mathrm{g \, cm^{-3}}$). See the text for a complete description.}
\label{fig:time_series}
\end{figure}
\end{center}

The gravitational signal generated by DCBHs is characterized by ${\cal D} \ll 1$, so there is a superposition of a discontinuous signal over a background noise: in this case, the component separation is easier to perform. Our Fig. \ref{fig:time_series} is a simple proof of concept of a signal processing method that allows to: (i) probe the presence of an underlying signal buried into a background noise and (ii) estimate its periodicity. The red solid line is the sum of $h_s(t)$, the signal from DCBHs collapse with ${\cal D}=0.2$ (this value has been chosen for an easier visualization on the plot), and the binary confusion background $h_n(t)$, as in \cite{Regimbau_2010} but with observations extended up to $z=15$. The signal is completely buried into the noise. Nonetheless, the blue points are the auto-correlation function of the signal, which shows a clear structure. It starts from a value $A_0 \sim 0.4$ and drops approximately after $\sim 4$ cycles of the signal to a value $\sim A_0/20$: this time separation is highlighted by the vertical orange line. In addition, the periodic behavior of the auto-correlation function resembles the intrinsic oscillations of the ringdown phase during the collapse. As a comparison, the green crosses represent the auto-correlation function for the noise component only: the absence of any structure is evident.
The two parameters which greatly affect the appearance of the auto-correlation function are the mass $M$ and the mean density $\rho$.
The total duration of the GWs emission depends on: (i) the duration of the collapse phase (proportional to $1/\sqrt{\rho}$) and (ii) the duration of the ringdown phase, which depends on $M$ (see the expression for $\nu_0$ in Sec. \ref{sec:theory}). In addition, the mass greatly affects the overall normalization (magnitude) of the signal, see Fig. \ref{fig:Comp_mass_z}. For masses $M > 10^5 \, \mathrm{\Msun}$, the signal would not be buried into the fore/background, then its detection would be straightforward. If we allow the mass of the population to vary below $M = 10^{5} \, \mathrm{\Msun}$ and the density to vary as well, what happens is that the auto-correlation function may have several rises and falls, around the typical time separations between these signals, as clearly shown in Fig. \ref{fig:time_series}.

This simple method proves the existence of an underlying periodic signal and provides a range of possible values for the period, i.e. the time separation needed for the auto-correlation function to drop. 
Anyway, a possible future detection of this signal with Ultimate-DECIGO would require the application of more sophisticated algorithms for data analysis, similar to those that have been proposed for the LISA experiment (e.g. \citealt{Crowder_2007}).

%%%%%%%%%%%%%%%%%%%%%%%%%%%%%%%%%%%%%%%%%%%%%%%%%%%%%%%%%%%%%%%%%%%%%%
%% SECTION 4: CONCLUSION
%%%%%%%%%%%%%%%%%%%%%%%%%%%%%%%%%%%%%%%%%%%%%%%%%%%%%%%%%%%%%%%%%%%%%%
\section{Conclusions}
\label{sec:conclusions}
In this work, we have employed modern waveforms and the improved knowledge on the DCBHs formation rate to estimate the gravitational signal emitted by the formation of DCBHs at $13 \lsim z \lsim 20$, in the so-called ``DCBHs Era" (see \citealt{Yue_2014}).
We have thoroughly discussed the reasons why the, unexpectedly high, DCBHs formation rate is not in tension with current estimates of the local BHs mass density. Indeed, there are currently no observational and/or theoretical indications ruling it out and further work is then required in this field, as also recent works (e.g. \citealt{Kormendy_2013, Comastri_2015}) suggest.

We have investigated the effects of a wide range of masses, spin parameters and mean mass densities on the gravitational signal, in order to build up a realistic ensemble of these high-redshift sources.
We have shown that, despite the very high rate of events for all-sky surveys ($\sim 10^4 \, \mathrm{yr^{-1}}$ up to $z\sim20$), the integrated signal from these sources is characterized by a very low duty-cycle (${\cal D}\sim 10^{-3}$), i.e. it is a shot-noise. This is a consequence of the small duration of the GW emission from this kind of sources ($\Delta T_{gw} \sim 2-30\, \mathrm{s}$).

Our results show that the gravitational signal lies above the foreseen sensitivity of the Ultimate-DECIGO observatory in the frequency range $(0.8-300) \, \mathrm{mHz}$, with a peak amplitude $\Omega_{gw} = 1.1 \times 10^{-54}$ at $\nu_{max} = 0.9 \, \mathrm{mHz}$ and a peak Signal-to-Noise Ratio of $\sim 22$ at $\nu = 20 \, \mathrm{mHz}$.
This amplitude is lower than the Galactic confusion noise in the same frequency band, generated by binary systems of compact objects.

For a gravitational signal characterized by ${\cal D} \ll 1$ and buried into this confusion noise, we have provided a very simple signal processing method to prove the existence of underlying periodic oscillations such as those we expect from DCBHs. However, more advanced techniques will be required to separate this signal from background and foreground noise components.
The signal investigated in this paper lies in the same frequency range of the one produced by NS binaries and, in the same way, acts as a noise for the cosmological background, which is the ultimate target of the experiment DECIGO. 
For this reason, it is very important to model any gravitational signal that may fall into the frequency range of interest for the primordial signal.

Despite all the technical difficulties, the actual observation of DCBHs may be a keystone in modern Cosmology, providing a significant contribution to the formation theory of Super-Massive Black Holes and to the understanding of the Cosmic Infrared Background.

%\section{Acknowledgements}
\vspace{+1cm}
FP thanks Raffaella Schneider, Stefania Salvadori, Bence Kocsis and Bin Yue for their very valuable comments and for useful discussions.
The research leading to these results has received funding from the European Research Council under the European Union's Seventh Framework Programme (FP/2007-2013)/ERC Grant Agreement n. 306476. 

%%%%%%%%%%%%%%%%%%%%%%%%%%%%%%%%%%%%%%%%%%%%%%%%%%%%%%%%%%%%%%%%%%%%%%
%%  REFERENCES
%%%%%%%%%%%%%%%%%%%%%%%%%%%%%%%%%%%%%%%%%%%%%%%%%%%%%%%%%%%%%%%%%%%%%% 

\end{document}